\documentclass[preprint,12pt]{article}
\usepackage[latin1]{inputenc}
\usepackage{mathrsfs,amsthm,mathtools,graphicx,epstopdf,color,verbatim,bm,bbm,amsmath,amsfonts,amssymb,newclude,nicefrac,amsfonts,varwidth,
graphicx,enumerate,hyperref,booktabs}
\usepackage{tabularx}
\usepackage{makecell} % new added
\usepackage{array} % new added
\usepackage{multirow} % new added
\usepackage{lipsum} % new added
\numberwithin{equation}{section}
\usepackage[top=1.1in, bottom=1.2in, left=0.8in, right=0.8in]{geometry}
\renewcommand{\thefootnote}{\ensuremath{\fnsymbol{footnote}}}

\newcommand{\dd}{\text{d}}
\newtheorem{theorem}{Theorem}[section]

\newtheorem{proposition}[theorem]{Proposition}

\begin{document}

\title{Efficient simulation of prices for European call options under Heston stochastic-local volatility model: a comparison of methods\footnotemark[1]}

\renewcommand{\thefootnote}{\fnsymbol{footnote}}

\author{
Meng Cai $^\text{a}$ and  
 Tianze Li $^\text{b}$ \footnotemark[2] \\
	\footnotesize $^\text{a}$ School of Statistics and Mathematics, Central University of Finance and Economics, Beijing, China\\
	\footnotesize $^\text{b}$ School of Economics, Central University of Finance and Economics, Beijing, China
}

\date{\today}

\maketitle

\footnotetext[1]{ This work was supported by the National Natural Science Foundation of China 
	(No. 12501581) and the Disciplinary Funding of Central University of Finance and Economics.}
\footnotetext[2]{ Corresponding author: litianze6@126.com}

\begin{abstract}

{\rm\small
         The Heston stochastic-local volatility  model,
         consisting of a asset price process and a  Cox--Ingersoll--Ross-type variance process,
          offers a wide range of applications in the financial industry.
           The pursuit for efficient model evaluation has been assiduously ongoing and  central to which is the numerical simulation of  
        CIR process. 
        Different from the weakly convergent noncentral chi-squared approximation used in \cite{van2014heston}, 
         this paper considers two  strongly convergent and positivity-preserving methods for CIR process under Lamperti transformation, namely, the truncated Euler method and the backward Euler method.
      It should be noted that these two methods are completely different.
        The explicit truncated Euler method is computationally effective and remains robust under high volatility,
         while the implicit backward Euler method provides high computational accuracy and stable performance.
         Numerical experiments on European call options are presented to show the superiority of different methods.
           } \\

%         \textbf{AMS subject classification:} {\rm\small 60H35, 60H15, 65C30}\\
%\textbf{PACS: 02.60.Lj}

\end{abstract}

\section{Introduction}\label{sec:introduction}

In the fundamental research of finance, precisely resolving the contradictions between pricing and risk management is of tremendous significance.
Recent developments on the stochastic-local volatility (SLV) model has paved the way for achieving a balance between perfect calibration and correct dynamics in financial engineering. 
This hybrid model synthesizes the features of both the stochastic volatility (SV) and local volatility (LV) models, thus benefiting from the main advantages of each.
The demand for stochastic volatility models (\cite{heston1993closed}, \cite{schobel1999stochastic})  stems from the observation that the volatilities of asset prices, when considered as time-series data, are inherently stochastic. 
Its advantages are twofold:
it can model the actual volatility as realistic as possible, so that the stock price distribution used in the model is close to the observed data; at the same time it fills in the gap in the implied volatility data set so that an appropriate volatility value can be used in derivative pricing and hedging.
However, the standard Heston model has only finitely many parameters, making it difficult to perfectly fit the implied volatility for all tenors and strike prices simultaneously. 
In particular, it typically demonstrates insufficient capability to fit the volatility smile of short-maturity options \cite{engelmann2011calibration}. 
An ``imperfect" calibration means that the model cannot even accurately reproduce today's market prices, which undermines the foundation of its pricing.
Local Volatility model (\cite{derman1998stochastic}, \cite{dupire1994pricing}) is a very powerful tool for market modeling, which generates arbitrage-free scenarios that are calibrated to all available options in the current market. 
However, a problem arose regarding the prediction of future volatility behavior. It usually results in a flattening forward volatility smile
 \cite{rebonato1999volatility}. 
More precisely, this model indicates that if the price of the underlying asset experiences significant changes in the future, the future implied volatility will tend to   be flat, which seriously contradicts the continuous pricing of tail risks in the options market. 
Such discrepancies may result in significant mispricing and suboptimal hedging strategies for path-sensitive derivatives, including  forward-starting options, step contracts, and snowball structures.
The SLV model, first introduced in \cite{JexHenWan}, is a powerful tool for both empirical studies and industry practice. 
It outperforms either the pure LV or SV models in calibrating European options, as well as in pricing and risk management (see, e.g., \cite{GuyHen2,lipton, renetal, Tianetal, van2014heston}).

In this paper we are concerned with the Heston stochastic-local volatility (HSLV) model,
where the volatility process follows Cox--Ingersoll--Ross (CIR) dynamics \cite{CIR}.
The CIR model describes interest movements as driven by a sole source of market risk,
which represents some macroeconomic external factors.
It has received considerable critical attention due to the useful properties, such as mean-reversion and non-negativity. 
In addition, the analytical tractability of the Heston model directly translates into the rapid calibration of its parameters.
In the past two decades, a number of researchers have attempted to conduct efficient evaluation for HSLV model.
The authors of pioneering works \cite{engelmann2011calibration} used a finite volume scheme  to calibrate the model after a suitable transformation and demonstrate its accuracy in numerical test cases using real market data.
Early work \cite{van2014heston} developed a nonparametric numerical scheme for efficient model evaluation.
Deep learning techniques could be used \cite{cuchiero2020generative},  allowing the calculation of model prices and model implied volatilities in an accurate way using only small sets of sample paths.
More recently, a Monte Carlo particle method is analyzed for the simulation of the calibrated HSLV model in \cite{reisinger2025numerical}.

The key difficulty in the model evaluation is the simulation of CIR process due to the presence of  unbounded diffusion term and  non-globally Lipschitz continuous coefficient.
Although there has been a few works \cite{CozMarRei,CozRei,DNS,hu2017backward},
it is still far from being well-understood.
It is well-known that when applying the Euler--Maruyama method, the variance process can become negative with nonzero probability.
A negative variance is meaningless both mathematically and financially, which can cause the entire simulation to collapse. 
Previous attempts in numerical simulations either take the maximum with 0 at each step, or use the non-central chi-square distribution.
However, the non-central chi-square approximation can only achieve convergence in a weak sense, namely, the laws of
the exact solution and the approximation almost coincide with each other.
The main purpose of this paper is to develop methods that are strongly convergent and can preserve the positivity.
A strongly convergent method makes sure that the corresponding trajectories of the exact solution and the approximation are close to each other.
Subsequently, it aims to deduce the improvement in overall option pricing so as to meet the requirements of different option applications.

The article is organized as follows. In the next section, we present the background and give a brief introduction to the underlying model.
We then introduce and analyse the numerical methods in Section \ref{sec:scheme}. 
The corresponding proofs are given in the
Appendix. 
Section \ref{sec;Numerical experiments} provides numerical experiment to the superiority of different methods.

\section{The considered model}\label{sec:preliminaries}

For a given time horizon $[0, T]$, we consider a complete filtered probability space $(\Omega, \mathcal{F}, \mathbb{P})$ with natural filtration $(\mathcal{F}_t)_{t \in [0, T]}$, which supports an $\mathcal{F}_t$--adapted two-dimensional standard Brownian motion $(W_t, \widetilde{W}_t)$. 
In this article, we focus on the HSLV model
\begin{equation}\label{eq:H-SLV}
	\begin{split}
		\left\{
		\begin{array}{lll} 
		  \dd S_t =r  S_t  \dd t +  
		  \sqrt{V_t}~{S_t}~\sigma(t,S_t)   \left( \rho  \dd W_t + \sqrt{ 1 - \rho^2}  \dd \widetilde{W}_t \right),~~S_0 >0, \\
			\dd V_t = \kappa ( \theta - V_t ) \dd t + \gamma ~ \sqrt{V_t} ~ \dd W_t,~~V_0 >0, \\
			\dd W_t ~\dd \widetilde{W}_t = \rho ~ \dd t,
		\end{array}\right.
	\end{split}
\end{equation}
where $S = \{S_t\}_{t\in [0,T]}$ describes the spot price of the underlying asset and $V = \{V_t\}_{t\in[0,T]}$ is a squared volatility process. 

It is common practice to follow two steps to calibrate HSLV model to market prices.
In the first step, we propose a suitable approach to solve the volatility $V = \{V_t\}_{t\in[0,T]}$.
The volatility process with CIR dynamics \cite{CIR} reads as
\begin{equation}\label{eq:cir}
	\dd V_t = \kappa ( \theta - V_t ) \dd t + \gamma ~ \sqrt{V_t} ~ \dd W_t,~~V_0 >0,
\end{equation}
where parameters $\kappa, \theta, \gamma > 0$ describing its mean-reversion rate, long-term mean, and volatility, respectively.
Let Feller's condition $ 2 \kappa \theta \ge \sigma^2$ hold. Then \eqref{eq:cir} admits a unique solution $X(t)$ satisfying
$$\mathbb{P} \{ V_t > 0, \forall~ t >0 \} =1$$
and 
$$\sup_{0 \leq t \leq T} \mathbb{E} \left[ | V_t |^p \right] \leq \infty, ~ p > - \frac{2 \kappa \theta}{\sigma^2}.$$
It benefits from the useful characteristics, including mean reversion and non-negativity \cite{AndPit}, 
and is well-received in the industry. 
Unfortunately, the exact solution of \eqref{eq:cir} is not provided explicitly. Therefore, an efficient and accurate numerical approximation is necessary.
In the second step, at each time grid, we need to calibrate $(\sigma(t,\cdot))$.
For the exact calibration of SV models, a consistency condition  to market prices is formulated in \cite{dupire1994pricing}. 
For general SLV models in \cite{JexHenWan}, 
\begin{equation}\label{Dupireiff}
	\sigma^{2}(t, s)=
	\frac{\sigma_{\text{Dup}}^{2}(t,s)}{\mathbb{E}[V_{t}|S_{t} = s]},
\end{equation}
where $\sigma_{\text{Dup}}$ is expressed by the Dupire formula, that is, for a given call option market prices $C(T,S)$ with maturity time $T$ and strike $S$
\begin{equation*}
	\sigma_{\text{Dup}}^{2}(T,S) = \left(\frac{\partial C(T,S)}{\partial T} + r S \frac{\partial C(T,S)}{\partial S} \right)
	\big/\left(\frac{S^2}{2}\cdot\frac{\partial^2 C(T,S)}{\partial S^2}\right).
\end{equation*}
Here we assume zero interest and dividend rates.

\section{The proposed numerical simulation}
\label{sec:scheme}
Fixed $T > 0$, $N \in \mathbb{N}_+$, step-size $\tau = \frac T N$, $t_n= n \tau, n =0,1, \cdots , N$.
For $m =0,1,2,\cdots,N-1$, we denote $\Delta W_{m} = W_{t_{m+1}} - W_{t_m}$ for simplicity.

\subsection{Simulation of volatility $V_t$}

To overcome the difficulties caused by  unbounded diffusion coefficient, we first apply Lamperti transformation $L_t = \sqrt{V_t}$ to \eqref{eq:cir} and  give an additive noise driven stochastic differential equation as follows, 
\begin{equation}\label{eq:cir2}
	 \dd L_t = \frac12 \kappa \left( \frac{\theta}{L_t} - L_t \right) \,\dd t +  \frac12 \sigma  \, \dd W_t, \quad  t \in (0, T],~~
			L_0 = \sqrt{V_0}.
\end{equation}
When applying the Euler--Maruyama method to \eqref{eq:cir2}, it may be negative with a positive probability.
Therefore, we will introduce two kinds of positivity-preserving numerical method in this subsection. 

\subsubsection{Backward Euler method}

The backward Euler method applied to \eqref{eq:cir2} gives 
\begin{equation}\label{eq:bem}
L_{t_{n+1}}^N = L_{t_n}^N +	\frac \kappa 2 \tau ( \frac{\theta}{L_{t_{n+1}}^N} - L_{t_{n+1}}^N )  +  \frac12 \sigma \Delta W_{n},
\end{equation}
which has a unique positive solution becuse of the Feller's condition.
It has been proved  in \cite{hu2025efficient} to be strongly convergent and stable. 
Nevertheless, this method is implicit. 
A nonlinear system need to be solved by iteration at each step, which may effect the efficiency of computation.

\subsubsection{Truncated Euler method}

Taking advantage of being explicit and easily implementable, a positivity-preserving scheme is proposed by virtue of truncating skill \cite{gao2022truncated}.
For any given $\tau \in (0,1]$, the truncated mapping $\pi_\tau : \mathbb{R} \to \mathbb{R}$ is defined by 
$$\pi_\tau(x) = ( b \tau^{\frac14})  \vee x , ~ \forall x \in \mathbb{R},$$
where $b \leq Y_0$ and is independent with the step-size $\tau$.
The so-called truncated Euler method is given as follows,
\begin{equation}\label{eq:truncated}
L_{t_{n+1}}^N = L_{t_n}^N + \frac \kappa 2 ~\tau \left( \frac{\theta}{\pi_{\tau}(L_{t_n}^N)} - \pi_{\tau}(L_{t_n}^N) \right) + \frac12 \sigma \Delta W_{n}.
\end{equation}
By taking $\overline{L}_{t_n}^N = \pi_{\tau} (L_{t_n}^N)$, we derive a positive numerical solution for \eqref{eq:cir2}.
We highlight that the truncated scheme with explicit time-stepping is more computationally efficient
than backward Euler method.

\begin{proposition}\label{prop:truncated}
Suppose $ 2 \kappa \theta \ge \sigma^2, T>0$. Then  the numerical solution generated by the truncated Euler method strongly converges to the exact solution of \eqref{eq:cir2}.

\begin{equation}
\sup_{1 \leq n \leq N} \left\| L_{t_n} - \overline{L}_{t_n}^N \right\|_{L^2(\Omega;\mathbb{R})} \leq 
C~\tau^{\frac 12}.
\end{equation}

\end{proposition}

Next, by inverse Lamperti transformation, we arrive at the following result.

\begin{theorem}
Under the assumption of Proposition \ref{prop:truncated}, 
let $\overline{V}_t^N = \left( \sum\limits_{k=1}^{\infty} \overline{L}_{t_k}^N \chi_{[t_{k-1},t_k)}(t) \right)^2$,
then for any $\tau \in (0,1]$, 
\begin{equation}
\sup_{1 \leq n \leq N} \mathbb{E} \left | V_{t_n}^N - \overline{V}_{t_n}^N \right | \leq C~\tau^{\frac 12}.
\end{equation}
\end{theorem}

\subsection{Simulation of asset price $S_t$}

For the asset price process,  we
follow \cite{broadie2006exact} and give the discretization of    $R_t:=\log (S_t)$ with $\hat{\sigma}(t,R_t):=\sigma(t,e^{R_t})$,  which reads
\begin{equation}\label{eq:asset-discret}
	\begin{split}
R_{t_{i+1}} & = R_{t_{i}} + \int_{t_i}^{t_{i+1}} \left( r - \frac12 \hat{\sigma}^2(s,R_s) V_s \right) \dd s   + \rho \int_{t_i}^{t_{i+1}} \hat{\sigma}(s,R_s) \sqrt{V_s} \dd W_s
 \\ & \quad + \sqrt{ 1 - \rho^2} \int_{t_i}^{t_{i+1}} \hat{\sigma}(s,R_s) \sqrt{V_s} \dd \widetilde{W}_s.
	\end{split}
\end{equation}
It follows from the volatility process that
\begin{equation}
	\int_{t_i}^{t_{i+1}} \sqrt{V_s} \dd W_s =
	\frac 1\gamma \left(V_{t_{i+1}} - V_{t_i} -\kappa \theta \tau + \kappa \int_{t_i}^{t_{i+1}}  V_s \dd s \right).
\end{equation}
Then,  the second integral in \eqref{eq:asset-discret} can be approximated as follows,
\begin{equation}
	\begin{split}
	\int_{t_i}^{t_{i+1}} & \hat{\sigma}(s,R_s) \sqrt{V_s} \dd W_s \approx
	\hat{\sigma}(t_i,R_{t_i}) \int_{t_i}^{t_{i+1}}  \sqrt{V_s} \dd W_s
\\ &	=\frac 1\gamma \hat{\sigma}(t_i,R_{t_i}) \left(V_{t_{i+1}} - V_{t_i} -\kappa \theta \tau + \kappa \int_{t_i}^{t_{i+1}}  V_s \dd s \right).
	\end{split}
\end{equation}
For the third integral in \eqref{eq:asset-discret}, using It\^o's isometry gives
\begin{equation}
	\int_{t_i}^{t_{i+1}} \hat{\sigma}(s,R_s) \sqrt{V_s} \dd \widetilde{W}_s \sim  \left( \int_{t_i}^{t_{i+1}} \hat{\sigma}^2(s,R_s) V_s \dd s \right)^{\frac12} \mathcal{N}(0,1),
\end{equation}
where $\mathcal{N}(0,1)$ obeys the standard normal distribution.
Furthermore, by the Euler discretization of all integrals with respect to time, we have
\begin{equation}
	\begin{split}
	R_{t_{i+1}} & \approx R_{t_{i}} + \int_{t_i}^{t_{i+1}} \left( r - \frac12 \hat{\sigma}^2(s,R_s) V_s \right) \dd s
		\\ & \quad +  \frac {1}{\gamma}~ \rho ~ \hat{\sigma}(t_i,R_{t_i}) \left(V_{t_{i+1}} - V_{t_i} -\kappa \theta \tau + \kappa \int_{t_i}^{t_{i+1}}  V_s \dd s \right)
		\\ & \quad + \sqrt{ 1 - \rho^2} \left( \int_{t_i}^{t_{i+1}} \hat{\sigma}^2(s,R_s) V_s \dd s \right)^{\frac12} \mathcal{N}(0,1)
		\\ & \approx R_{t_i} + r \tau - \frac12 \tau \hat{\sigma}^2({t_i},R_{t_i}) V_{t_i}   
		+  \frac {1}{\gamma} ~ \rho ~ \hat{\sigma}({t_i},R_{t_i}) \left(V_{t_{i+1}} - V_{t_i} -\kappa \theta \tau + \kappa \tau  V_{t_i}  \right)
		\\ & \quad + \sqrt{ (1 - \rho^2) \tau \hat{\sigma}^2({t_i},R_{t_i}) V_{t_i} } ~ \mathcal{N}(0,1).
	\end{split}
\end{equation}

\subsection{Simulation of $\sigma^2(t,S)$}

The discrete scheme becomes
\begin{equation}
	\begin{split}
	R_{i+1,j} & = R_{i,j} + r \tau - \frac12 \tau \hat{\sigma}^2(t_i,R_{i,j})V_{i,j} + \sqrt{1- \rho^2} \cdot \sqrt{\hat{\sigma}^2(t_i,R_{i,j})V_{i,j}
		\tau}\cdot \mathcal{N}(0,1)
	\\ & \quad +  \frac {1}{\gamma} \rho\hat{\sigma}(t_i,R_{i,j}) \left(V_{i+1,j} - V_{i,j} -\kappa \theta \tau + \kappa \tau V_{i,j}  \right).
	\end{split}
\end{equation}
For the discretization of volatility, we employ the following approximation
\begin{equation}
  \hat{\sigma}^2(t_i,R_{i,j}) :=    \sigma^2(t_i,e^{R_{i,j}}) 
  = \frac{\sigma_{LV}^2(t_i,S_{i,j})}
  {\mathbb{E} \left[ V_{t_i }| S_{t_i} = S_{i,j} \right]},
\end{equation}
where
\begin{equation}
	\sigma_{LV}^2(t_i,S_{i,j}) = \frac{\frac{\partial C(t,S)}{\partial t} + r S \frac{\partial C(t_i,S)}{\partial  S} }{ \frac 12 S^2 \frac{\partial^2 C(t_i,S)}{\partial S^2} } \Big|_{S=S_{i,j},t=t_i}.
\end{equation}
The finite difference method is adopted to compute local volatility. 
To mitigate deviations caused by excessively small second-order derivative results, a lower bound is implemented in the calculation to enhance the stability.

\section{Numerical experiments}\label{sec;Numerical experiments}
We perform some numerical experiments in this section to visually confirm our theoretical conclusion.
The codes were written by the Python programming language and all the tests were performed by Spyder on a YOGA Pro 14s ASP9 laptop with an AMD Ryzen CPU and 16 GB of RAM in a Windows 11 environment.

In our model setup, we primarily conduct numerical experiments on the HSLV model, using the standard Monte Carlo approach. The market environment is based on the classical Heston model, with the following parameters: $\gamma = 0.95$ (vol-of-vol), $\kappa = 1.05$ (mean reversion rate), $\rho = -0.315$ (correlation), $\theta = 0.0855$ (long-run variance), $S_0 = 1$, $r = 0$, $v_0 = 0.0945$ (initial variance), and $T = 5$.
For HSLV model, we introduce a modification parameter $p = 0.25$ and 
define continuous adjusted parameters  as follows: $\bar{\gamma} = (1 - p)  \gamma,~\bar{\kappa} = (1 + p) \kappa,~ \bar{\rho} = (1 + p)  \rho,~\bar{\theta} = (1 - p) \theta,~\bar{v}_0 = (1 + p)  v_0,~ r = 0,~S_0 = 1,~T = 5$.

Firstly, we test the ability of the truncated and backward Euler methods applied to compute conditional expectations. 
This ensures that the binning approach could be reasonably applied to the calculation of volatility, which is consistent with the description in \cite{andreasen2011volatility}. 
In the simulation of conditional expectations, we used the 2D-COS method \cite{ruijter2012two} as the benchmark and chose 20 bins, 10,000 paths, and a step-size of 0.001.

\begin{figure}[!htb]
	\centering
	\begin{varwidth}[t]{\textwidth}
		\includegraphics[width=2.8in]{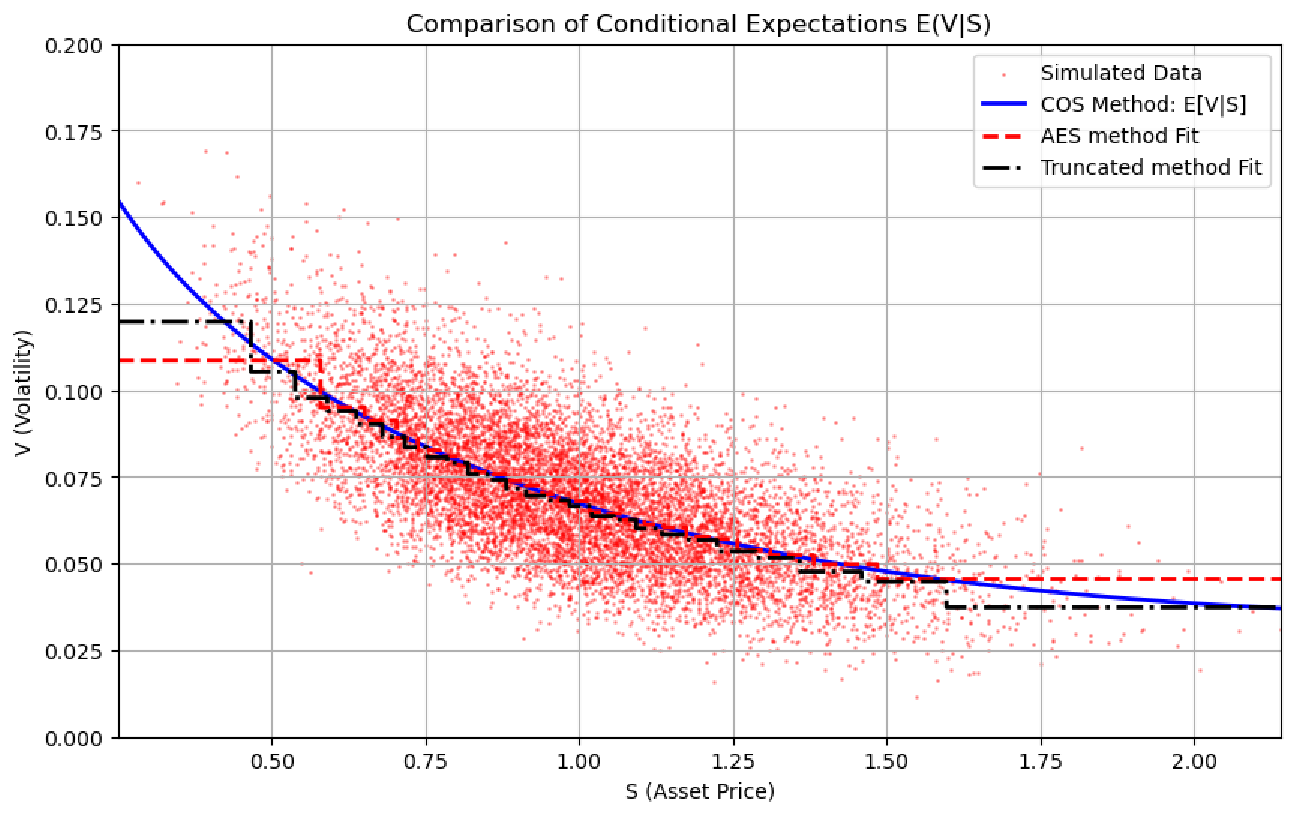}
		\qquad
		\includegraphics[width=2.8in]{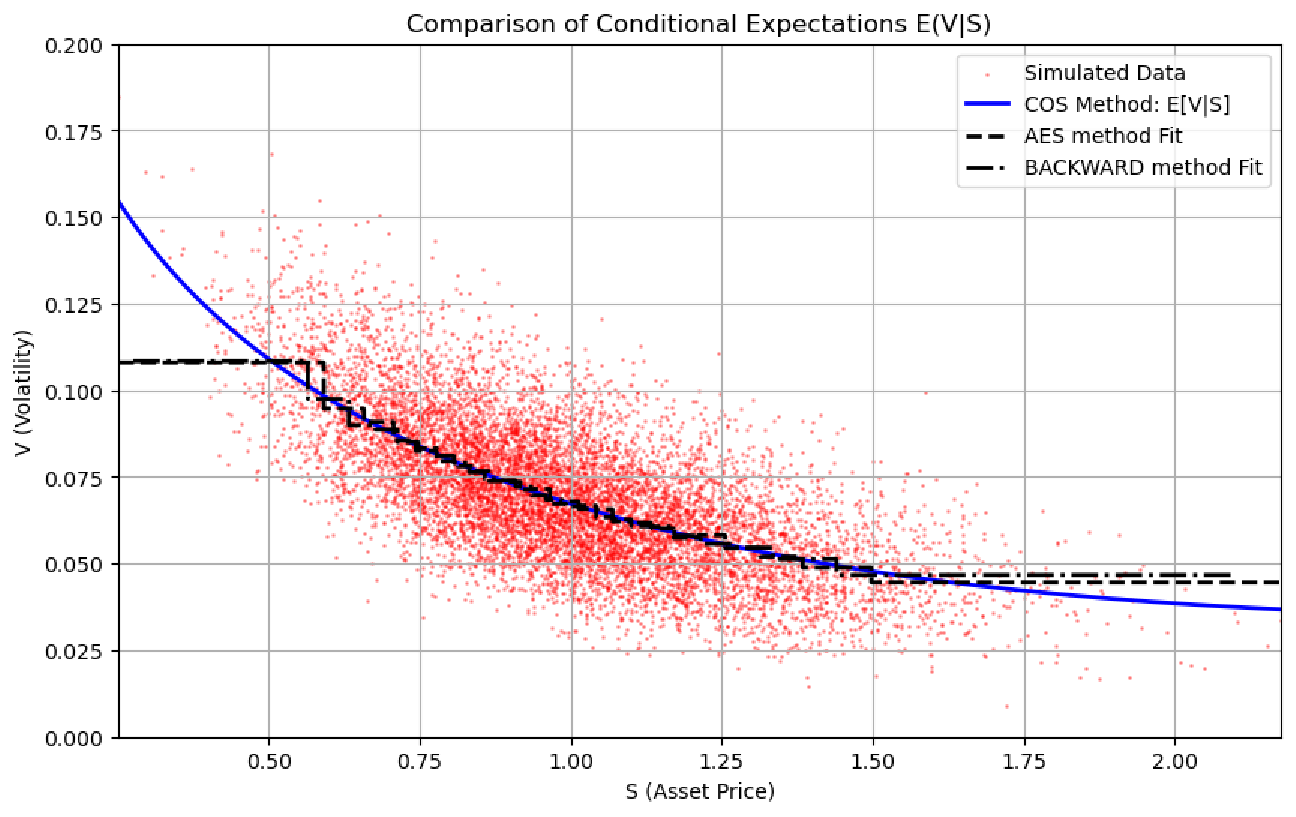}
	\end{varwidth}
	\caption{ Comparison of expectations}
	\label{F1}
\end{figure}

From Figure \ref{F1}, we can observe that both the truncated Euler method and the backward Euler method effectively adopt the binning approach to estimate the conditional expectation. 
This guarantees the validity of our subsequent experiments. 
To more clearly highlight the differences, we multiplied the error by a factor of 100 in the numerical experiments, and the results are presented in the following tables.

\begin{table}[!htb]
	\centering
	\tabcolsep = 0.13cm
	\renewcommand{\arraystretch}{1.15}
	\caption{\label{1} The error of four different methods}
	\vspace{0.5em}	
	\begin{tabularx}{0.8\textwidth}{cccccc}
		\toprule[0.5mm]
		\ \ \multirow{2}{*}{$N$} & \multicolumn{4}{l}{\makecell[c]{\qquad \qquad \qquad $K= 70 \%$ }} \\
		\cmidrule(lr){3-6} & & $\ \ \ Euler\ \ $ & $\ \ AES\ \ $ & $\ \ \ Truncated\ \ $ & $\ \ \ Backward\ \ $\\
		\midrule[0.5mm]
		\ \ $5$  & \quad & $16.55 \% (0.30\%) ~~$ & $17.56 \% (0.47\%) ~~$ & $13.49 \% (0.24\%) ~~$ & $15.49 \% (0.28\%) ~~$ \\
		\ \ $10$  & \quad & $13.60 \% (0.31\%) ~~$ & $14.92 \% (0.40\%) ~~$ & $12.14 \% (0.35\%) ~~$ & $12.20 \% (0.33\%) ~~$ \\
		\ \ $25$  & \quad & $13.10 \% (0.23\%) ~~$ & $14.32 \% (0.48\%) ~~$ & $13.11 \% (0.23\%) ~~$ & $11.51 \% (0.33\%) ~~$ \\
		\ \ $40$ & \quad & $1.88 \% (0.97\%) ~~$ & $4.13 \% (0.62\%) ~~$ & $4.66 \% (0.53\%) ~~$ & $0.06 \% (0.66\%) ~~$ \\
		\bottomrule[0.5mm]
	\end{tabularx}
\end{table}

\begin{table}[!htb]
	\centering
	\tabcolsep = 0.13cm
	\renewcommand{\arraystretch}{1.15}
	\caption{\label{2} The error of four methods}
	\vspace{0.5em}	
	\begin{tabularx}{0.8\textwidth}{cccccc}
		\toprule[0.5mm]
		\ \ \multirow{2}{*}{$N$} & \multicolumn{4}{l}{\makecell[c]{\qquad \qquad \qquad $K= 100 \%$ }} \\
		\cmidrule(lr){3-6} & & $\ \ \ Euler\ \ $ & $\ \ AES\ \ $ & $\ \ \ Truncated\ \ $ & $\ \ \ Backward\ \ $\\
		\midrule[0.5mm]
		\ \ $5$  & \quad & $15.79 \% (0.16\%) ~~$ & $18.65 \% (0.31\%) ~~$ & $14.08 \% (0.24\%) ~~$ & $15.97 \% (0.10\%) ~~$ \\
		\ \ $10$  & \quad & $12.40 \% (0.20\%) ~~$ & $15.94 \% (0.39\%) ~~$ & $12.71 \% (0.25\%) ~~$ & $12.66 \% (0.20\%) ~~$ \\
		\ \ $25$  & \quad & $11.40 \% (0.23\%) ~~$ & $14.22 \% (0.30\%) ~~$ & $13.07 \% (0.23\%) ~~$ & $11.49 \% (0.23\%) ~~$ \\
		\ \ $40$ & \quad & $3.61 \% (1.26\%) ~~$ & $3.54 \% (0.70\%) ~~$ & $3.86 \% (0.74\%) ~~$ & $0.46 \% (0.68\%) ~~$ \\
		\bottomrule[0.5mm]
	\end{tabularx}
\end{table}

\begin{table}[!htb]
	\centering
	\tabcolsep = 0.13cm
	\renewcommand{\arraystretch}{1.15}
	\caption{\label{3} The error of four methods}
	\vspace{0.5em}	
	\begin{tabularx}{0.8\textwidth}{cccccc}
		\toprule[0.5mm]
		\ \ \multirow{2}{*}{$N$} & \multicolumn{4}{l}{\makecell[c]{\qquad \qquad \qquad $K= 150 \%$ }} \\
		\cmidrule(lr){3-6} & & $\ \ \ Euler\ \ $ & $\ \ AES\ \ $ & $\ \ \ Truncated\ \ $ & $\ \ \ Backward\ \ $\\
		\midrule[0.5mm]
		\ \ $5$  & \quad & $14.72 \% (0.50\%) ~~$ & $17.13 \% (1.77\%) ~~$ & $13.96 \% (0.85\%) ~~$ & $17.32 \% (1.58\%) ~~$ \\
		\ \ $10$  & \quad & $12.11 \% (0.34\%) ~~$ & $13.69 \% (0.92\%) ~~$ & $13.44 \% (0.76\%) ~~$ & $14.35 \% (0.61\%) ~~$ \\
		\ \ $25$  & \quad & $11.05 \% (0.41\%) ~~$ & $12.19 \% (0.48\%) ~~$ & $13.48 \% (0.52\%) ~~$ & $13.11 \% (0.45\%) ~~$ \\
		\ \ $40$ & \quad & $3.85 \% (1.50\%) ~~$ & $2.87 \% (1.15\%) ~~$ & $4.80 \% (1.35\%) ~~$ & $1.44 \% (1.15\%) ~~$ \\
		\bottomrule[0.5mm]
	\end{tabularx}
\end{table}

When the number of steps is small, the truncated Euler method usually has the lowest error. 
However, as the number of steps grows, this method becomes less sensitive to changes in step-size. 
When the number of steps is large, the backward Euler method is more accurate than the others. 
Nevertheless, in terms of runtime, our statistics (for example, with 40 steps: the AES method takes 0.1114 seconds, the Euler method takes 0.0939 seconds, the truncated Euler method takes 0.0733 seconds, and the backward Euler method takes 137.9476 seconds) indicate that the backward Euler method requires more computational cost than the truncated Euler method. 
The reason behind this may be that a nonlinear system needs to be solved by a Newton--Raphson iteration at each step.
In summary, the two methods discussed in this paper both demonstrate clear advantages, and their applicability depends on the specific context and computational requirements.

\begin{figure}[!htb]
	\centering
	\begin{varwidth}[t]{\textwidth}
		\includegraphics[width=2.2in]{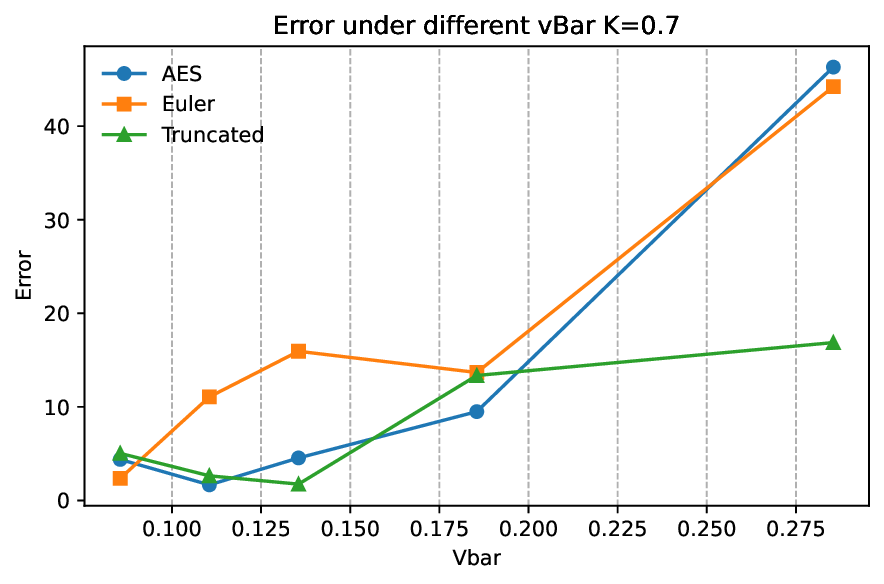}
		~
		\includegraphics[width=2.2in]{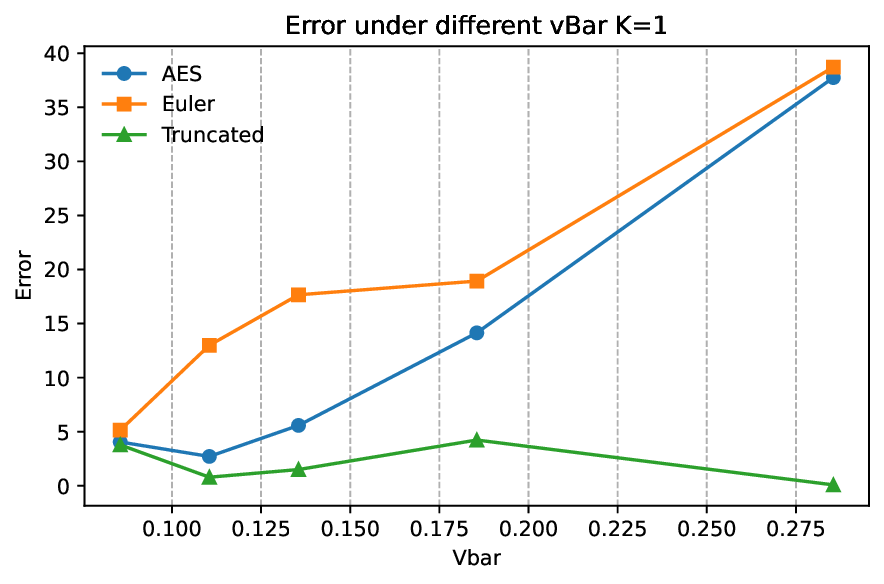}
		~
		\includegraphics[width=2.2in]{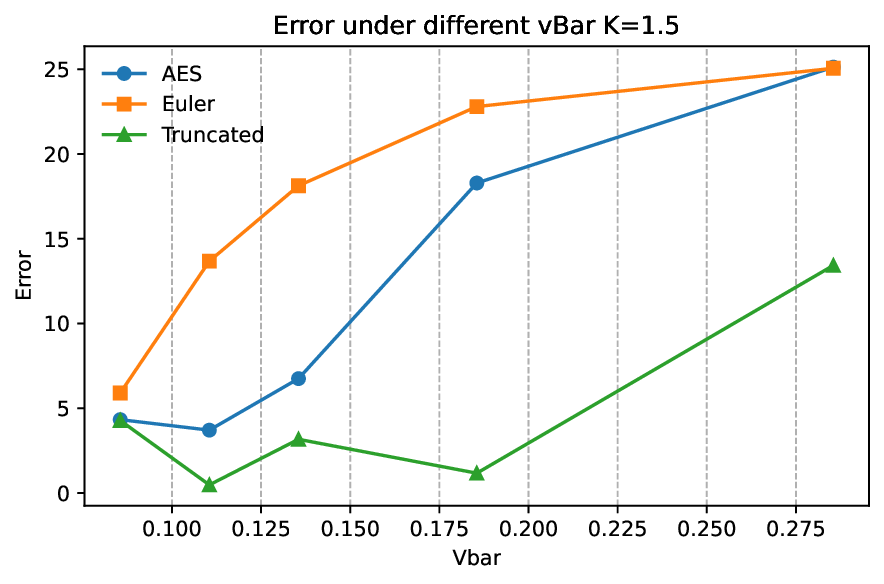}
	\end{varwidth}
	\caption{Error under different parameters $\theta$}
	\label{F2}
\end{figure}

Secondly, we evaluate the numerical performance of the AES method, the Euler method, and the truncated Euler method  under different values of the volatility parameter $\theta$. 
From Figure \ref{F2}, we can infer that as $\theta$ increases, the absolute errors  of both the AES method and the Euler method increase significantly. 
In contrast, the truncated Euler method exhibits remarkable robustness in error control and its error growth rate remains lower.
In financial terms, increasing the parameter $\theta$ is equivalent to simulating an extreme scenario where volatility experiences a sharp surge and market stress intensifies. 
For example, this is similar to the sudden spikes in volatility witnessed during financial crises.
In such a case, the distribution of the underlying asset paths exhibits heavier tails and higher kurtosis, imposing stringent requirements on the numerical accuracy of option-pricing models.
The AES method is highly efficient under normal market conditions, especially, when $\theta$ is moderate. 
Nevertheless, it is extremely sensitive to parameter changes. 
When $\theta$ increases, the discrete errors are magnified, indicating that the AES method is not suitable for stress testing and tail-risk modeling. Under extreme market conditions, it may significantly underestimate or overestimate option values, thus introducing substantial model risk.
In contrast, the truncated Euler method shows strong numerical stability. 
As $\theta$ increases, its error only increases slightly, which indicates that it has lower sensitivity and better adaptability to the average magnitude of the volatility process.
This robustness stems from its design. Through the application of appropriate mathematical transformations, the method can captures the complex geometric characteristics of distribution tails more effectively, ensuring that the  approximation of integration maintains high accuracy even under extreme parameterizations.

\begin{figure}[!htb]
	\centering
	\begin{varwidth}[t]{\textwidth}
		\includegraphics[width=2.2in]{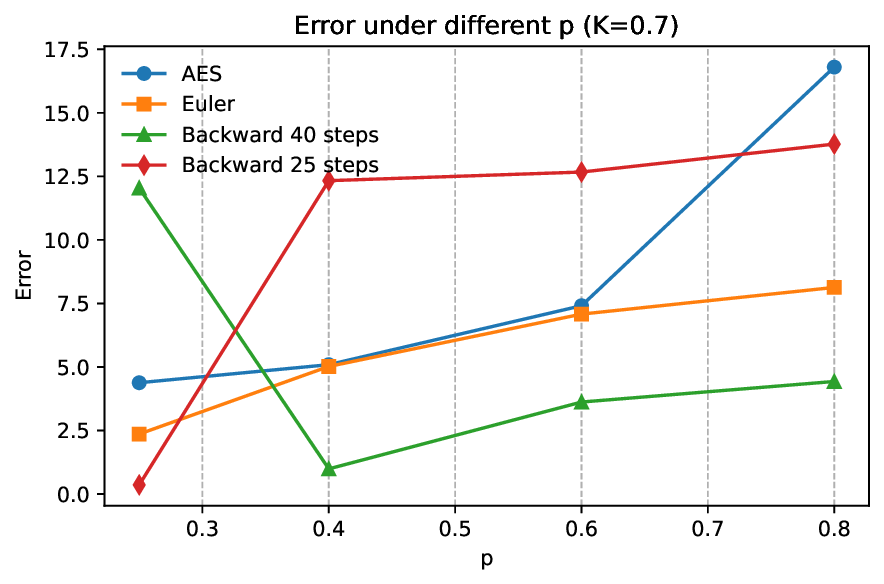}
		~
		\includegraphics[width=2.2in]{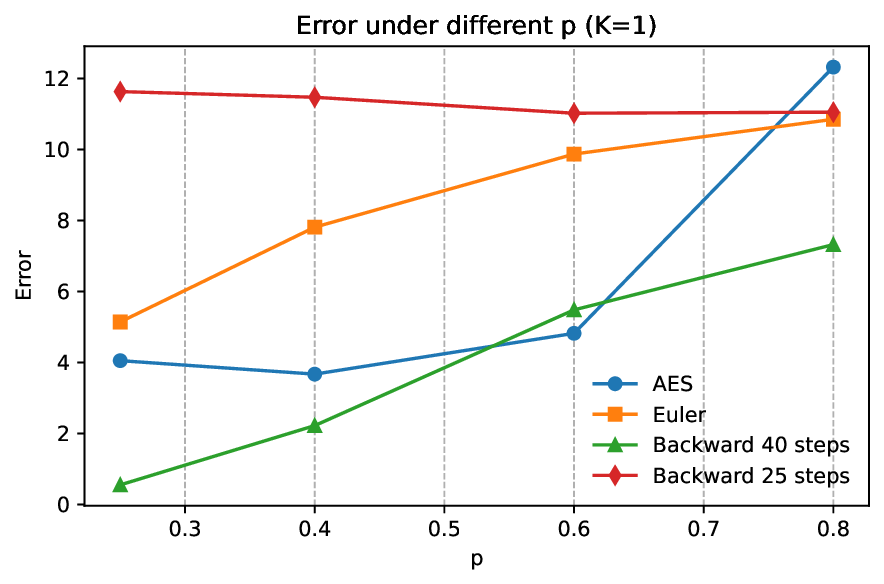}
		~
		\includegraphics[width=2.2in]{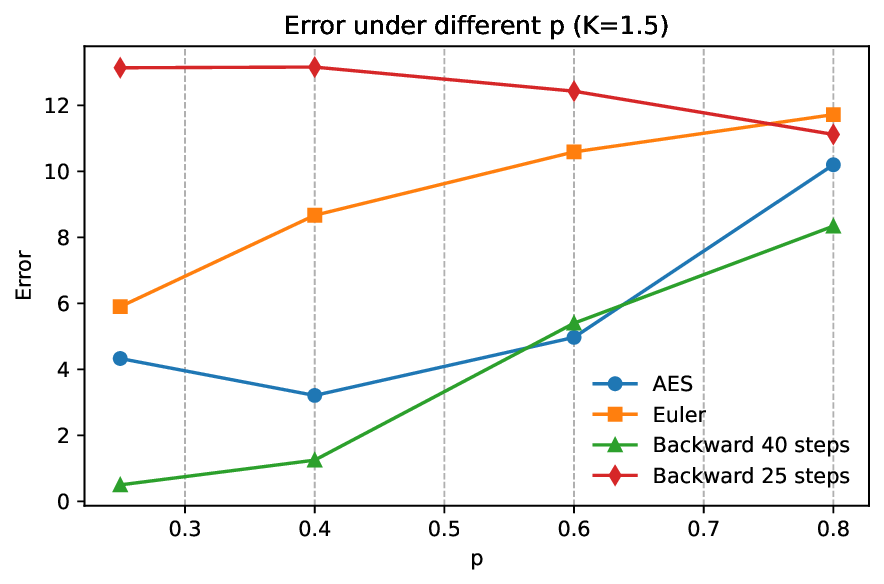}
	\end{varwidth}
	\caption{Error under different parameters $p$}
	\label{F3}
\end{figure}

Finally, we examine the numerical performance of the AES method, the Euler method, and the backward method  under different values of the parameter $p$. The results shown in Figure \ref{F3} indicate that as $p$ increases, 
the backward Euler method consistently performs better than the other methods in terms of calibration errors.
Nonetheless, we also observed that the backward method is quite sensitive to the step-size. 
When the number of steps increases from 25 to 40, its calibration accuracy fluctuates significantly. 
This suggests that careful attention must be paid to the choice of step-sizes when applying the backward Euler method. 
If the priority is computational speed, the number of steps for the backward Euler method should be decreased. 
This allows for a quick estimation of risk exposures within an acceptable error tolerance, enabling practitioners to seize short-lived trading opportunities or implement timely hedging strategies. 
On the other hand, if accuracy is the main concern, increasing the number of steps will produce more precise and reliable asset valuation results in model validation. 
These results are crucial for profit-and-loss analysis.

\section{Conclusion}
In this study, we focused on the efficient evaluation of  HSLV model, especially CIR variance process.
When applying traditional Euler--Maruyama method, the variance process can become negative with nonzero
probability.
The non-central chi-square approximation can be positivity-preserving, but is typically weak convergent, 
restricting the practical application in option pricing.
In this paper, we introduce truncated Euler method and backward Euler method, which are both positivity-preserving and strongly convergent. 
Some experiments on European call options are carried out to compare different methods.

Numerical experiments indicate that the truncated Euler method is computationally efficient and remains robust against parameter changes. This is particularly evident when volatility spikes and tail risks become more pronounced. This property makes it an excellent choice in scenarios where both efficiency and reliability are of critical importance.
The backward Euler method also demonstrates a remarkable characteristic.  
It showcases excellent accuracy and the most stable performance in error growth under various stress scenarios. 
However, it incurs a significantly higher computational cost because solving implicit equations demands iterative method, such as the 
Newton-Raphson method.

This gives rise to an important practical concern: how to select the step-size? 
Smaller step-sizes improve accuracy but also lengthen the runtime. 
Hence, it is essential to strike a balance between precision and efficiency based on the specific application. Overall, our results show that both proposed methods have distinct advantages. 
The truncated Euler method offers a practical solution for large-scale simulations and  real-time risk management, where computational resources and execution speed are crucial. 
On the contrary, the backward Euler method is more appropriate for stress testing, model validation, and scenarios where high accuracy is extremely important, such as profit-and-loss attribution. 
Together, these methods enrich the numerical tools available for HSLV model simulation, providing flexible strategies for practitioners with different objectives in option pricing and risk management.

\section{Appendix: strong convergence of truncated Euler method for CIR model}

{\em Proof of Proposition \ref{prop:truncated}: }
For simplicity, we denote $\phi(x) = \frac \kappa2 \left( \frac \theta x - x \right)$. It is easy to check that
\begin{equation}
\left | \phi(x) - \phi (y) \right | \leq \frac{2 | \kappa |  \vee \kappa \theta}{4} \left( 1 + \frac 1{x^2} +  \frac 1{y^2} \right) | x -y |,
\end{equation}
and
\begin{equation}
(x-y) \left( \phi(x) - \phi (y) \right ) \leq  - \frac{\kappa}{2} | x -y |^2.
\end{equation}
Furthermore,
\begin{equation}\label{eq:phi-lip}
\left | \phi(\pi_{\tau}(x)) - \phi(\pi_{\tau}(y)) \right |^2 \leq C ~\tau^{-1}  | x -y |^2,
\end{equation}
and
\begin{equation}\label{eq:phi-onelip}
(x-y) \left( \phi(\pi_{\tau}(x)) - \phi(\pi_{\tau}(y)) \right ) \leq  - \frac{\kappa}{2} | x -y |^2.
\end{equation}
Next, we continue the proof in two steps.

{\bf Step 1: Estimate of $\| L_{t_n}^N - \overline{L}_{t_n}^N \|_{L^2(\Omega; \mathbb{R})}.$}

Let $e_n=  L_{t_n}^N - \overline{L}_{t_n}^N $, then $e_0 =0$ and 
 \begin{equation}
 \begin{split}
 e_{n+1} - e_n & = \int_{t_n}^{t_{n+1}} \phi(L_s) \dd s - \phi (\pi_{\tau}(\overline{L}_{t_n}^N)) \tau
 \\ &
 = \int_{t_n}^{t_{n+1}} \left[ \phi(L_s) - \phi (\pi_{\tau}(L_{t_n}))\right] \dd s 
 - \tau \left[\phi (\pi_{\tau}(L_{t_n})) - \phi (\pi_{\tau}(\overline{L}_{t_n}^N))  \right].
 \end{split}
 \end{equation}
With the aid of Young's Inequality, \eqref{eq:phi-lip} and \eqref{eq:phi-onelip}, we get
\begin{equation}
\begin{split}
| e_{n+1} |^2 &\leq | e_n |^2 + 2 \left |  \int_{t_n}^{t_{n+1}} \left[ \phi(L_s) - \phi (\pi_{\tau}(L_{t_n}))\right] \dd s \right |^2
+ 2 \tau^2 \left | \phi (\pi_{\tau}(L_{t_n})) - \phi (\pi_{\tau}(\overline{L}_{t_n}^N))) \right |^2
\\ & \quad + 2 e_n \int_{t_n}^{t_{n+1}} \left[ \phi(L_s) - \phi (\pi_{\tau}(L_{t_n}))\right] \dd s + 2 \tau  e_n \left[\phi (\pi_{\tau}(L_{t_n})) - \phi (\pi_{\tau}(\overline{L}_{t_n}^N)))  \right]
\\& \leq ( 1 + \tau )  | e_n |^2 + \left(2 + \frac{1}{\tau}\right) \left |  \int_{t_n}^{t_{n+1}} \left[ \phi(L_s) - \phi (\pi_{\tau}(L_{t_n}))\right] \dd s \right |^2
\\ & \quad + 2 \tau^2  \left | \phi (\pi_{\tau}(L_{t_n})) - \phi (\pi_{\tau}(\overline{L}_{t_n}^N)))  \right |^2 + 2 \tau  e_n \left[\phi (\pi_{\tau}(L_{t_n})) - \phi (\pi_{\tau}(\overline{L}_{t_n}^N)))  \right]
\\ & \leq ( 1 + C \tau )  | e_n |^2 + \left(2 + \frac{1}{\tau}\right) \left |  \int_{t_n}^{t_{n+1}} \left[ \phi(L_s) - \phi (\pi_{\tau}(Y(t_n)))\right] \dd s \right |^2.
\end{split}
\end{equation}
By iteration, we have 
\begin{equation}
| e_{n+1} |^2 \leq  C \left( \tau \sum_{i=0}^n | e_i |^2 + \frac 1\tau \sum_{i=0}^n \left |  \int_{t_i}^{t_{i+1}} \left[ \phi(L_s) - \phi (\pi_{\tau}(L_{t_i}))\right] \dd s \right |^2 \right).
\end{equation} 
Taking expectation on both sides and applying Gronwall's inequality give 
\begin{equation}
\mathbb{E} \left[ | e_{n} |^2 \right] \leq  C  \frac 1\tau \sum_{i=0}^n \mathbb{E} \left[ \left |  \int_{t_i}^{t_{i+1}} \left[ \phi(L_s) - \phi (\pi_{\tau}(L_{t_i}))\right] \dd s \right |^2 \right].
\end{equation} 
In the sequence, we bound $\mathbb{E} \left[ \left |  \int_{t_i}^{t_{i+1}} \left[ \phi(L_{s}) - \phi (\pi_{\tau}(L_{t_i}))\right] \dd s \right |^2 \right]$.
 Thanks to the fundamental inequality, we derive
 \begin{equation}
 \begin{split}
 \mathbb{E} & \left[ \left |  \int_{t_i}^{t_{i+1}} \left[ \phi(L_{s}) - \phi (\pi_{\tau}(L_{t_i}))\right] \dd s \right |^2 \right]
\\ & \leq 2 \mathbb{E} \left[ \left |  \int_{t_i}^{t_{i+1}} \left[ \phi(L_{s}) - \phi (L_{t_i})\right] \dd s \right |^2 \right]
 + 2 \tau^2 \mathbb{E}  \left[ \left |  \phi(L_{t_i}) - \phi (\pi_{\tau}(L_{t_i}))\right |^2 \right].
 \end{split}
 \end{equation}
On one hand, it follows by It\^{o} isometry and moment bounds of $Y(t)$ that
  \begin{equation}
 \begin{split}
  \mathbb{E}& \left[ \left |  \int_{t_i}^{t_{i+1}} \left[ \phi(L_s) - \phi (L_{t_i})\right] \dd s \right |^2 \right]
  = \mathbb{E} \left[ \left |  \int_{t_i}^{t_{i+1}}  \int_{t_i}^{s} \dd \phi(L_r) \dd s \right |^2 \right]
  \\ & \quad = \mathbb{E} \left[ \left | \int_{t_i}^{t_{i+1}} (t_{i+1} - r ) \phi' (L_r) \phi (L_r) \dd r + \int_{t_i}^{t_{i+1}} \frac \sigma 2  (t_{i+1} - r ) \phi' (L_r) \dd W_r \right |^2 \right]
  \\ & \quad \leq C \left ( \mathbb{E} \left[ \left | \int_{t_i}^{t_{i+1}} (t_{i+1} - r ) \phi' (L_r) \phi (L_r) \dd r \right |^2 \right] 
  + \mathbb{E} \int_{t_i}^{t_{i+1}}   \left | (t_{i+1} - r ) \phi' (L_r) \right |^2  \dd   r \right )
   \\ & \quad \leq C ~ \tau^4 + C ~ \tau^3.
  \end{split}
 \end{equation}
On the other hand, note that
\begin{equation}\label{eq:1}
\begin{split}
\left| \phi(L_{t_i}) - \phi (\pi_{\tau}(L_{t_i})) \right|^2 &\leq 
 C \left( 1 + \frac 1{| L_{t_i} |^4} + \frac 1{| \pi_{\tau} (L_{t_i}) |^4} \right)
 \left | L_{t_i} - \pi_{\tau} (L_{t_i}) \right|^2
 \\ & \leq C \left( 1 + \frac 1{| L_{t_i} |^4}  \right)
 \left | L_{t_i} - b \tau^{\frac14} \right|^2
 \chi_{\{ L_{t_i} < b \tau^{\frac14}\}}
 \\ & \leq C \tau^{\frac12}  \left( 1 + | L_{t_i} |^{-4}  \right) 
 \chi_{\{ L_{t_i} < b \tau^{\frac14}\}}.
\end{split}
\end{equation}
Using H\"older's inequality and Chebyshev's inequality leads to
\begin{equation}\label{eq:2}
\begin{split}
2 \tau^2 & \mathbb{E} \left[ \left| \phi(L_{t_i}) - \phi (\pi_{\tau}(L_{t_i})) \right|^2 \right] \leq  C \tau^{\frac52} 
 \left( \mathbb{P}\{ L_{t_i} < b \tau^{\frac14}\} 
 + \mathbb{E} \left[ | L_{t_i} |^{-4}  \chi_{\{ L_{t_i} < b \tau^{\frac14}\} } \right] \right) 
 \\ & \leq C \tau^{\frac52}  \left( \mathbb{P}\left\{ L_{t_i}^{-1} > \left( b \tau^{\frac14} \right)^{-1} \right\} 
 + \mathbb{E} \left[ | L_{t_i} |^{-6}  \right]^{\frac 23} \left[ \mathbb{P}\left\{ L_{t_i}^{-1} > \left( b \tau^{\frac14} \right)^{-1} \right\}  \right]^{\frac 13}  \right) 
 \\ & \leq C ~ \tau^3.
\end{split}
\end{equation}
Hence,
\begin{equation}
\mathbb{E} \left[ | e_{n} |^2 \right] \leq  C  \frac 1\tau \sum_{i=0}^n \tau^3 \leq C ~ \tau.
\end{equation}

{\bf Step 2: Estimate of $\| L_{t_n} - \overline{L}_{t_n}^N \|_{L^2(\Omega; \mathbb{R})}.$}

Introduce an auxiliary process $\overline{L}_{t_n} = \pi_{\tau}(Y_{t_n})$, then
\begin{equation}
\mathbb{E} \left[ \left| \overline{L}_{t_n}^N - L_{t_n} \right|^2 \right]
 \leq 2 \mathbb{E} \left[ \left| \overline{L}_{t_n}^N - \overline{L}_{t_n} \right|^2 \right] + 2 \mathbb{E} \left[ \left| \overline{L}_{t_n} - L_{t_n} \right|^2 \right].
\end{equation}
 On one hand, by the property of $\pi_{\tau}$, we have
 \begin{equation}
 \mathbb{E} \left[ \left| \overline{L}_{t_n}^N - \overline{L}_{t_n} \right|^2 \right]
  \leq \mathbb{E} \left[ \left| L_{t_n}^N - L_{t_n} \right|^2 \right] \leq C \tau.
 \end{equation}
 On the other hand, following the similar manner of the treatment of \eqref{eq:1} and \eqref{eq:2}, and using Chebyshev's inequality arrives at 
\begin{equation}
\mathbb{E} \left[ \left| \overline{L}_{t_n} - L_{t_n} \right|^2 \right] \leq C \tau^2.
\end{equation}
Therefore, we finish the proof. \qquad
\qquad \qquad \qquad \qquad \qquad \qquad \qquad \qquad \qquad  \qquad \qquad \qquad \qquad $\square$

\end{document}